\documentclass[12pt]{article}%
\usepackage{amssymb}
\usepackage{amsfonts}
\usepackage{amsmath}
\usepackage{graphicx}%
\providecommand{\U}[1]{\protect\rule{.1in}{.1in}}

\begin{document}

\title{{\LARGE Quantum and pseudoclassical descriptions of nonrelativistic spinning
particles in noncommutative space }}
\author{{\large T. C. Adorno}\thanks{adorno@dfn.if.usp.br}{\large , M. C.
Baldiotti}\thanks{baldiott@fma.if.usp.br} \ {\large and D. M. Gitman}%
\thanks{gitman@dfn.if.usp.br}\\\textit{Instituto de F\'{\i}sica, Universidade de S\~{a}o Paulo,}\\\textit{Caixa Postal 66318, CEP 05508-090, S\~{a}o Paulo, S.P., Brazil}}
\maketitle

\begin{abstract}
We construct a nonrelativistic wave equation for spinning particles in the
noncommutative space (in a sense, a $\theta$-modification of the Pauli
equation). To this end, we consider the nonrelativistic limit of the $\theta
$-modified Dirac equation. To complete the consideration, we present a
pseudoclassical model (\textit{\`{a} la} Berezin-Marinov) for the
corresponding nonrelativistic particle in the noncommutative space. To justify
the latter model, we demonstrate that its quantization leads to the $\theta
$-modified Pauli equation. We extract $\theta$-modified interaction between a
nonrelativistic spin and a magnetic field from such a Pauli equation and
construct a $\theta$-modification of the Heisenberg model for two coupled
spins placed in an external magnetic field. In the framework of such a model,
we calculate the probability transition between two orthogonal EPR
(Einstein-Podolsky-Rosen) states for a pair of spins in an oscillatory
magnetic field and show that some of such transitions, which are forbidden in
the commutative space, are possible due to the space noncommutativity. This
allows us to estimate an upper bound on the noncommtativity parameter.

\end{abstract}

\section{Introduction}

It is known that QFT's in noncommutative spaces induce the so-called $\theta
$-modified relativistic wave equations, which, at present, are interpreted as
Schr\"{o}dinger equations of relativistic quantum mechanics in noncommutative
spaces, see e.g.,
\cite{BouBen05,GitKup08,ChaGanSa04,HarKumKha04,KokOkaSai04,SchChaGanHaz05,Smail02,Delduc}%
. Calculations in the framework of such theories allow one to evaluate the
influence of space noncommutativity on different physical effects and, e.g.,
to establish upper bonds on the noncommutativity parameters $\theta$, see
\cite{AdoBalChaGitTur09,MocPosRoi00,ChaJab01,ChaJabTur01}. In the present
article, we construct a nonrelativistic wave equation for spinning particles
in the noncommutative space (in a sense, a $\theta$-modification of the Pauli
equation). In course of solving the above mentioned problem, we consider the
nonrelativistic limit of the $\theta$-modified Dirac equation. It should be
noted that the form of the latter equation depend essentially on the point of
view how to write the action of the spinor field in an external background in
the noncommutative space. We consider two possible actions, one obtained from
simple Moyal modification \cite{Moyal} (introducing Moyal products into the
ordinary Dirac field action
\cite{BouBen05,GitKup08,MinRaaSei99,Hayakawa00,Szabo,MatSusTou00,ArfYav00,RiaJab00,ChaJab01,ArdSad01}%
) and another action obtained by the so-called Seiberg-Witten (SW) map, see
\cite{SeiWit99,DouNek01,GurJacPiPol01,Cai01,CarHarKosLanOka01,BicGriPopSchWul01}%
.

It should be noted that the noncommutativity can be also justified by a group
theoretical analysis of the Galilean symmetry in nonrelativistic particle
systems. A relation between spin and noncommutativity can be studied on
examples of classical models of nonrelativistic spinning particles, considered
in \cite{LukStiZak97,DuvHor00}, \cite{Horvathy03}\ and \cite{HorPly02}. The
corresponding quantum versions of the models describe anyons and their
interaction with external fields in the relativistic and nonrelativistic
context \cite{HorPly04,HorPly05}.

Doing nonrelativistic limits in both $\theta$-modified Dirac equations, we
follow the standard approach, see e.g. \cite{FolWou50,Schwe}, separating
\textquotedblleft big\textquotedblright\ and \textquotedblleft
small\textquotedblright\ two-component spinors in the Dirac four-component
wave function. The equation for the big spinor is just the nonrelativistic
wave equation for spinning particles. To complete the consideration, we
present a pseudoclassical model (\textit{\`{a} la} Berezin-Marinov
\cite{BerMar77,All1}) for the corresponding nonrelativistic particle in the
noncommutative space. To justify the model, we demonstrate that its
quantization leads to the $\theta$-modified wave equation for nonrelativistic
spinning particle.

Then, we consider one of possible applications of the obtained general result.
Namely, we extract $\theta$-modified interaction between a nonrelativistic
spin and a magnetic field from the $\theta$-modified Pauli equation. With such
an interaction in hands, we construct a $\theta$-modification of the
Heisenberg model for two coupled spins placed in an external magnetic field.
In the framework of such a model, we calculate the probability transition
between two orthogonal EPR (Einstein-Podolsky-Rosen) states for a pair of
spins in an oscillatory magnetic field and show that some of such transitions,
which are forbidden in the commutative space, are possible due to the space noncommutativity.

\section{Nonrelativistic limit of $\theta$-modified Dirac equation}

\subsection{Simple Moyal modification}

Here, we construct an action of the spinor field $\Psi$ (in an external
electromagnetic background $A^{\mu}$) in a noncommutative (NC) space,
introducing the Moyal star product in the ordinary spinor field action (we
call this a simple Moyal modification in what follows). In such a way, we
derive the following action\footnote{By bold, we denote three-vectors, e.g.,
$\mathbf{a}=\left(  a^{i}=-a_{i},\ i=1,2,3\right)  .$}%
\begin{align}
&  S_{\mathrm{M}}^{\theta}=\int dx\mathcal{L}_{\mathrm{M}}^{\theta
},\ \ \ \mathcal{L}_{\mathrm{M}}^{\theta}=\bar{\Psi}\left(  x\right)
\star\left(  \gamma^{\mu}\hat{P}_{\mu}-mc\right)  \star\Psi\left(  x\right)
\,,\nonumber\\
&  \hat{P}_{\mu}\left(  x\right)  =\hat{p}_{\mu}-\frac{e}{c}A_{\mu}\left(
x\right)  ,\ \ \hat{p}_{\mu}=i\hslash\partial_{\mu},\ \ A^{\mu}=\left(
A^{0},\mathbf{A}\right)  ,\ \ \mathbf{A}=\left(  A^{i},\ i=1,2,3\right)  \,,
\label{wm0}%
\end{align}
where the Moyal star product \textquotedblleft$\star$\textquotedblright\ is
defined as%
\[
f\left(  x\right)  \star g\left(  x\right)  =f\left(  x\right)  e^{\frac{i}%
{2}\overleftarrow{\partial}_{\mu}\theta^{\mu\nu}\vec{\partial}_{\nu}}g\left(
x\right)  \,,
\]
with $f\left(  x\right)  $ and $g\left(  x\right)  $ being arbitrary
functions, see e.g.
\cite{BouBen05,GitKup08,Szabo,MatSusTou00,ArfYav00,RiaJab00,ChaJab01,ArdSad01}%
, and $x=\left(  x^{0}=ct,\,x^{i},\ \ i=1,2,3\right)  $. Here and in what
follows, the subindex \textrm{M }reminds us that we use the simple Moyal modification.

Then the $\theta$-modified Dirac equation with an external electromagnetic
field, for a particle with the charge $e$ (for an electron $e=-\left\vert
e\right\vert $) and the mass $m$ is identified with Euler-Lagrange equation
$\delta S_{\mathrm{M}}^{\theta}/\delta\bar{\Psi}=0$. Thus, we obtain:%
\begin{align}
&  \left(  \gamma^{\mu}\hat{P}_{\mu}^{\theta}-mc\right)  \Psi\left(  x\right)
=0\,,\nonumber\\
&  \hat{P}_{\mu}^{\theta}=\hat{p}_{\mu}-\frac{e}{c}A_{\mu}\left(  x^{\mu
}+\frac{i}{2}\theta^{\mu\nu}\partial_{\nu}\right)  \,, \label{wm1}%
\end{align}
where $\gamma^{\mu}=\left(  \gamma^{0},\boldsymbol{\gamma}\right)  $ are Dirac
gamma-matrices, see, e.g. \cite{BouBen05,GitKup08}.

As was mentioned above, our aim is to derive a nonrelativistic quantum
mechanical description of a spinning particle in a noncommutative space. Since
in the nonrelativistic case, time and the configuration space are considered
separately, it is consistent (and natural) to treat space and space-time
noncommutativity effects separately. In what follows, we consider the case of
the space noncommutativity only, which implies $\theta^{0\mu}=0$. Such a
choice is also supported by the fact that in the framework of noncommutative
classical or quantum mechanics, there exist physical motivations just for the
space noncommutativity. In particular, the parameters $\theta^{ij}$ admit many
close analogies with a constant magnetic field both from the algebraic and
dynamical points of view \cite{Smail02,Delduc}.

We can rewrite (\ref{wm1}) in the Hamiltonian form (see, e.g.
\cite{BouBen05,GitKup08,AdoBalChaGitTur09}),%
\begin{align}
&  i\hslash\partial_{t}\Psi\left(  x\right)  =\widehat{\mathbb{H}}%
_{\mathrm{D}}\left(  \hat{q}\mathbf{,\hat{p}}\right)  \Psi\left(  x\right)
,\ \ \partial_{t}=\frac{\partial}{\partial t}\,,\nonumber\\
&  \widehat{\mathbb{H}}_{\mathrm{D}}\left(  \hat{q}\mathbf{,\hat{p}}\right)
=c\boldsymbol{\alpha}\cdot\mathbf{\hat{P}}\left(  \hat{q}\right)
+mc^{2}\gamma^{0}+eA^{0}\left(  \hat{q}\right)  \,, \label{e1}%
\end{align}
where $\Psi\left(  x\right)  $, is a bispinor, $\mathbf{\hat{P}}\left(
\hat{q}\right)  =\mathbf{\hat{p}}-\frac{e}{c}\mathbf{A}\left(  \hat{q}\right)
$, $\hat{q}=\hat{q}^{\mu}=\left(  x^{0},\hat{q}^{i}\right)  $,%
\begin{align}
&  \hat{q}^{i}=x^{i}-\frac{1}{2\hslash}\theta^{ij}\hat{p}^{j},\ \ \hat{p}%
^{j}=-i\hslash\partial_{j}\,,\nonumber\\
&  \left[  \hat{q}^{i},\hat{q}^{j}\right]  =i\theta^{ij},\ \ \left[  \hat
{q}^{i},\hat{p}^{j}\right]  =i\hslash\delta^{ij}\,,\ \ \left[  \hat{p}%
^{i},\hat{p}^{j}\right]  =0\,. \label{c1}%
\end{align}

In the first order in $\theta$, equation (\ref{wm1}) reads:%
\begin{equation}
\left\{  \gamma^{\mu}\left(  \hat{P}_{\mu}\left(  x\right)  -\frac{ie}%
{2c}\partial_{\alpha}A_{\mu}\left(  x\right)  \theta^{\alpha\beta}%
\partial_{\beta}\right)  -mc\right\}  \Psi\left(  x\right)  =0\,. \label{wm2}%
\end{equation}
Being written in the Hamiltonian form, equation (\ref{wm2}) reads:%
\begin{align}
&  i\hslash\partial_{t}\Psi\left(  x\right)  =\hat{\mathbb{H}}_{\mathrm{M}%
}^{\theta}\Psi\left(  x\right)  \,,\ \hat{\mathbb{H}}_{\mathrm{M}}^{\theta
}=\hat{\mathbb{H}}_{\mathrm{D}}+\Delta\hat{\mathbb{H}}_{\mathrm{M}}^{\theta
}\,,\nonumber\\
&  \hat{\mathbb{H}}_{\mathrm{D}}=c\boldsymbol{\alpha}\cdot\mathbf{\hat{P}%
}+eA_{0}+mc^{2}\gamma^{0}\,,\nonumber\\
&  \Delta\hat{\mathbb{H}}_{\mathrm{M}}^{\theta}=\frac{e}{2\hslash}\left[
\boldsymbol{\nabla}\left(  \boldsymbol{\alpha}\cdot\mathbf{A}-A_{0}\right)
\times\mathbf{\hat{p}}\right]  \cdot\boldsymbol{\theta}\,,\nonumber\\
&  \mathbf{\hat{P}}=\mathbf{\hat{p}}-\frac{e}{c}\mathbf{A}%
\,,\ \ \boldsymbol{\theta}=\left(  \theta^{i}=\frac{1}{2}\varepsilon
_{ijk}\theta^{jk}\right)  \,,\ \ \boldsymbol{\alpha}=\gamma^{0}%
\boldsymbol{\gamma}\,. \label{wm3}%
\end{align}

Let us consider the nonrelativistic limit of the latter equation following the
standard scheme and doing transformations \textit{\`{a} la}\ Foldy-Wouthuysen,
see e.g. \cite{FolWou50,Schwe}. The Hamiltonian $\hat{\mathbb{H}}_{\mathrm{M}%
}^{\theta}$ (\ref{wm3}) is written in terms of an odd operator\footnote{An
operator which has only matrix elements connecting the upper and lower
components of the Dirac spinor\ is classified as odd $\mathcal{O}$ and an
operator which does not have such elements is classified as even $\mathcal{E}%
$.} $\mathcal{O}_{\mathrm{M}}=c\boldsymbol{\alpha}\cdot\mathbf{\hat{P}%
}+\left(  e/2\hslash\right)  \left[  \boldsymbol{\nabla}\left(
\boldsymbol{\alpha}\cdot\mathbf{A}\right)  \times\mathbf{\hat{p}}\right]
\cdot\boldsymbol{\theta}$\ and an even operator $\mathcal{E}_{\mathrm{M}%
}=mc^{2}\gamma^{0}+eA_{0}-\left(  e/2\hslash\right)  \left[
\boldsymbol{\nabla}A_{0}\times\mathbf{\hat{p}}\right]  \cdot\boldsymbol{\theta
}$. First, we perform the canonical transformation $\Psi^{\left(  1\right)
}=e^{i\hat{S}_{\mathrm{M}}^{\left(  1\right)  }}\Psi$, with $\hat
{S}_{\mathrm{M}}^{\left(  1\right)  }=\left(  2imc^{2}\right)  ^{-1}\gamma
^{0}\mathcal{O}_{\mathrm{M}}$, trying to eliminate odd operators from
$\hat{\mathbb{H}}_{\mathrm{M}}^{\theta}$\ (deriving the nonrelativistic
approximation, we neglected terms of the order $O\left(  \left(
mc^{2}\right)  ^{-3}\right)  $ independent of $\theta$),%
\begin{equation}
i\hslash\partial_{t}\Psi^{\left(  1\right)  }=\hat{\mathbb{H}}_{\mathrm{M}%
}^{\theta\left(  1\right)  }\Psi^{\left(  1\right)  },\ \ \hat{\mathbb{H}%
}_{\mathrm{M}}^{\theta\left(  1\right)  }=e^{i\hat{S}_{\mathrm{M}}^{\left(
1\right)  }}\left(  \hat{\mathbb{H}}_{\mathrm{M}}^{\theta}-i\hslash
\partial_{t}\right)  e^{-i\hat{S}_{\mathrm{M}}^{\left(  1\right)  }}\,.
\label{wm4}%
\end{equation}
For $\hat{\mathbb{H}}_{\mathrm{M}}^{\theta\left(  1\right)  }$, we obtain%
\begin{align}
&  \hat{\mathbb{H}}_{\mathrm{M}}^{\theta\left(  1\right)  }=\mathcal{E}%
_{\mathrm{M}}+\mathcal{E}_{\mathrm{M}}^{\left(  1\right)  }+\mathcal{O}%
_{\mathrm{M}}^{\left(  1\right)  }+\mathcal{O}_{\mathrm{M}}^{\prime\left(
1\right)  },\ \ \mathcal{E}_{\mathrm{M}}^{\left(  1\right)  }=\frac{1}%
{2mc^{2}}\gamma^{0}\mathcal{O}_{\mathrm{M}}^{2}\nonumber\\
&  -\frac{1}{8m^{2}c^{4}}\left[  \mathcal{O}_{\mathrm{M}},\left(  e\left[
\mathcal{O}_{\mathrm{M}},\left(  A_{0}-\frac{1}{2\hslash}\left[
\boldsymbol{\nabla}A_{0}\times\mathbf{\hat{p}}\right]  \cdot\boldsymbol{\theta
}\right)  \right]  +i\hslash\partial_{t}\mathcal{O}_{\mathrm{M}}\right)
\right]  \,,\nonumber\\
&  \mathcal{O}_{\mathrm{M}}^{\left(  1\right)  }=\frac{\gamma^{0}}{2mc^{2}%
}\left(  e\left[  \mathcal{O}_{\mathrm{M}},A_{0}\right]  +i\hslash\partial
_{t}\mathcal{O}_{\mathrm{M}}-\frac{e}{2\hslash}\varepsilon_{ijk}\left[
\mathcal{O}_{\mathrm{M}},\left(  \partial_{i}A_{0}\right)  \hat{p}^{j}\right]
\theta^{k}\right)  \,,\nonumber\\
&  \mathcal{O}_{\mathrm{M}}^{\prime\left(  1\right)  }=-\frac{1}{3m^{2}c^{4}%
}\mathcal{O}_{\mathrm{M}}^{3}\,. \label{wm5}%
\end{align}

The first canonical transformation does not eliminate odd operators of the
order $\left(  mc^{2}\right)  ^{-1}$, that is why we need to perform a second
canonical transformation with the generator $\hat{S}_{\mathrm{M}}^{\left(
2\right)  }=\left(  2imc^{2}\right)  ^{-1}\gamma^{0}\mathcal{O}_{\mathrm{M}%
}^{\left(  1\right)  }$. Thus, we obtain a Schr\"{o}dinger equation with the
Hamiltonian%
\begin{align}
&  \hat{\mathbb{H}}_{\mathrm{M}}^{\theta\left(  2\right)  }=e^{i\hat
{S}_{\mathrm{M}}^{\left(  2\right)  }}\left(  \hat{\mathbb{H}}_{\mathrm{M}%
}^{\theta\left(  1\right)  }-i\hslash\partial_{t}\right)  e^{-i\hat
{S}_{\mathrm{M}}^{\left(  2\right)  }}=\mathcal{E}_{\mathrm{M}}+\mathcal{E}%
_{\mathrm{M}}^{\left(  1\right)  }+\mathcal{O}_{\mathrm{M}}^{\left(  2\right)
}\,,\nonumber\\
&  \mathcal{O}_{\mathrm{M}}^{\left(  2\right)  }=\frac{e}{2mc^{2}}\gamma
^{0}\left(  \left[  \mathcal{O}_{\mathrm{M}}^{\left(  1\right)  }%
,A_{0}\right]  -\frac{1}{2\hslash}\varepsilon_{ijk}\left[  \mathcal{O}%
_{\mathrm{M}}^{\left(  1\right)  },\left(  \partial_{i}A_{0}\right)  \hat
{p}^{j}\right]  \theta^{k}\right)  +\mathcal{O}_{\mathrm{M}}^{\prime\left(
1\right)  }-\hslash\partial_{t}\hat{S}_{\mathrm{M}}^{\left(  2\right)  }\,.
\label{wm6}%
\end{align}

The operator $\hat{\mathbb{H}}_{\mathrm{M}}^{\theta\left(  2\right)  }$ still
contains the odd operator $\mathcal{O}_{\mathrm{M}}^{\left(  2\right)  }$ of
the order $\left(  mc^{2}\right)  ^{-2}$, the latter operator can be
eliminated by a third canonical transformation with the generator $\hat
{S}_{\mathrm{M}}^{\left(  3\right)  }=\left(  2imc^{2}\right)  ^{-1}\gamma
^{0}\mathcal{O}_{\mathrm{M}}^{\left(  2\right)  }$. Finally, we get the
Schr\"{o}dinger equation with the Hamiltonian $\hat{\mathbb{H}}_{\mathrm{M}%
}^{\theta\left(  3\right)  }$ and the wave function $\Psi^{\left(  3\right)
}.$ The new Hamiltonian $\hat{\mathbb{H}}_{\mathrm{M}}^{\theta\left(
3\right)  }$ is already an even operator, it has the form%
\begin{equation}
\hat{\mathbb{H}}_{\mathrm{M}}^{\theta\left(  3\right)  }=e^{i\hat
{S}_{\mathrm{M}}^{\left(  3\right)  }}\left(  \hat{\mathbb{H}}_{\mathrm{M}%
}^{\theta\left(  2\right)  }-i\hslash\partial_{t}\right)  e^{-i\hat
{S}_{\mathrm{M}}^{\left(  3\right)  }}=\mathcal{E}_{\mathrm{M}}+\mathcal{E}%
_{\mathrm{M}}^{\left(  1\right)  }=\mathrm{diag}\left(  mc^{2}+\hat
{H}_{\mathrm{M}}^{\theta},\hat{h}^{\theta}\right)  , \label{wm7.2}%
\end{equation}
and
\[
\Psi^{\left(  3\right)  }=e^{i\hat{S}_{\mathrm{M}}^{\left(  3\right)  }%
}e^{i\hat{S}_{\mathrm{M}}^{\left(  2\right)  }}e^{i\hat{S}_{\mathrm{M}%
}^{\left(  1\right)  }}\Psi=\left(
\begin{array}
[c]{c}%
\psi\left(  x\right) \\
\chi\left(  x\right)  \sim\left(  mc^{2}\right)  ^{-1}\psi\left(  x\right)
\end{array}
\right)  .
\]

In the approximation under consideration, equations for upper
\textquotedblleft big\textquotedblright\ bispinor $\psi$ and inferior
\textquotedblleft small\textquotedblright\ bispinor $\chi$ are independent. We
interpret $\psi$ as the wave functions of the nonrelativistic spinning
particle with the Hamiltonian $\hat{H}_{\mathrm{M}}^{\theta}$.

Retaining only terms of the order $\left(  mc^{2}\right)  ^{-1}$ in $\hat
{H}_{\mathrm{M}}^{\theta}$ (\ref{wm7.2}), we obtain the following equation for
$\psi:$
\begin{align}
&  i\hslash\partial_{t}\psi\left(  x\right)  =\hat{H}_{\mathrm{M}}^{\theta
}\psi\left(  x\right)  \,,\nonumber\\
&  \hat{H}_{\mathrm{M}}^{\theta}=\frac{1}{2mc^{2}}\mathcal{O}_{\mathrm{M}}%
^{2}+eA_{0}-\frac{e}{2\hslash}\left[  \boldsymbol{\nabla}A_{0}\times
\mathbf{\hat{p}}\right]  \cdot\boldsymbol{\theta}\,, \label{wm8}%
\end{align}
where%
\[
\mathcal{O}_{\mathrm{M}}=c\boldsymbol{\alpha}\cdot\mathbf{\hat{P}}+\frac
{e}{2\hslash}\left[  \boldsymbol{\nabla}\left(  \boldsymbol{\alpha}%
\cdot\mathbf{A}\right)  \times\mathbf{\hat{p}}\right]  \cdot\boldsymbol{\theta
}\ .
\]
The complete Hamiltonian $\hat{H}_{\mathrm{M}}^{\theta}$ (with all the terms
up to the order $\left(  mc^{2}\right)  ^{-2}$) are represented in the Appendix.

Taking as $\mathbf{A}$ a vector potential that corresponds to a homogeneous
external magnetic field $\mathbf{B}=\left(  B^{i}\left(  t\right)  \right)  $
(in the symmetric gauge)%
\begin{equation}
A^{i}\left(  x\right)  =\frac{1}{2}\varepsilon_{ijk}B^{j}\left(  t\right)
x^{k}\,,\label{wm9.1}%
\end{equation}
where $\varepsilon_{ijk}$ is the Levi-Civita symbol in three dimensions with
the normalization $\varepsilon_{123}=1$, we obtain the following Hamiltonian:%
\begin{align}
&  \hat{H}_{\mathrm{M-Nonrel}}^{\theta}=\frac{1}{2m}\mathbf{\hat{P}}%
^{2}+eA_{0}-\frac{e}{2\hslash}\left(  \left[  \boldsymbol{\nabla}A_{0}%
\times\mathbf{\hat{p}}\right]  \cdot\boldsymbol{\theta}\right)  \nonumber\\
&  +\frac{e}{4m\hslash c}\left(  \mathbf{\hat{P}}\cdot\left[  \mathbf{B}%
\times\left[  \mathbf{\hat{p}}\times\boldsymbol{\theta}\right]  \right]
\right)  -\boldsymbol{\hat{\mu}}\cdot\mathbf{B}\left(  1+\frac{e}{4\hslash
c}\left(  \mathbf{B}\cdot\boldsymbol{\theta}\right)  \right)  \,,\label{wm10}%
\end{align}
where $\boldsymbol{\hat{\mu}}$ is particle magnetic moment related to the spin
operator $\mathbf{\hat{s}}$ as follows:
\[
\boldsymbol{\hat{\mu}}=\frac{e}{mc}\mathbf{\hat{s}}=\mu_{B}\boldsymbol{\sigma
}\,,\ \ \mu_{B}=\frac{e\hslash}{2mc}\,,\ \ \mathbf{\hat{s}}=\frac{\hslash}%
{2}\boldsymbol{\sigma}\,,\ \ \boldsymbol{\sigma}=\left(  \sigma^{i}\right)
\,.
\]

The equation (\ref{wm10}) (as well as eq. (\ref{wm8})) is not gauge invariant,
since the Dirac equation (\ref{wm3})\ is not gauge invariant. That is why, we
cannot interpret the factor in front of the operator $\boldsymbol{\hat{\mu}}$
as a physical magnetic field, this factor is gauge dependent. For example,
choosing the Landau gauge $A^{1}=-By,\ A^{2}=0$, this factor is reduced to
$\mathbf{B}$ and does not depend on $\theta$ at all\textbf{.}

\subsection{$\theta$-modified pseudoclassical action for nonrelativistic
spinning particle}

Considering the nonrelativistic limit of the equation (\ref{e1}), we obtain:%
\begin{align}
&  i\hslash\partial_{t}\psi\left(  x\right)  =\hat{H}_{\mathrm{M-Nonrel}%
}^{\theta}\psi\left(  x\right)  \,,\nonumber\\
&  \hat{H}_{\mathrm{M-Nonrel}}^{\theta}=\frac{1}{2m}\mathbf{\hat{P}}%
^{2}\left(  \hat{q}\right)  +eA^{0}\left(  \hat{q}\right)  -\boldsymbol{\hat
{\mu}}\cdot\mathbf{B}^{\theta}\,,\nonumber\\
&  \hat{\mathcal{B}}^{i}=-\frac{\varepsilon_{ijk}}{2}F_{jk}\left(  \hat
{q}\right)  =-\frac{\varepsilon_{ijk}}{2}\left\{  F_{jk}\left(  x\right)
+\frac{ie}{\hslash c}\left[  A_{j}\left(  \hat{q}\right)  ,A_{k}\left(
\hat{q}\right)  \right]  \right\}  \,, \label{e2}%
\end{align}
where%
\begin{align*}
F_{\mu\nu}\left(  x\right)   &  =\frac{ic}{\hslash e}\left[  \hat{P}_{\mu
}\left(  x\right)  ,\hat{P}_{\nu}\left(  x\right)  \right]  \Longrightarrow\\
F_{\mu\nu}\left(  \hat{q}\right)   &  =\partial_{\mu}A_{\nu}\left(  \hat
{q}\right)  -\partial_{\nu}A_{\mu}\left(  \hat{q}\right)  +\frac{ie}{\hslash
c}\left[  A_{\mu}\left(  \hat{q}\right)  ,A_{\nu}\left(  \hat{q}\right)
\right]  \,.
\end{align*}

If the external magnetic field is homogeneous, with potentials (\ref{wm9.1}),
it follows from (\ref{e2}) that $\mathbf{B}^{\theta}$ does not depend on
spacial\ coordinates, is linear in $\theta$, and has the form\footnote{It
should be noted that in the first order in $\theta,$ the Hamiltonian
(\ref{e2}) is reduced to the one (\ref{wm10}).}:%
\begin{equation}
\mathbf{B}^{\theta}=\left[  1+\frac{e}{4\hslash c}\left(  \mathbf{B}%
\cdot\boldsymbol{\theta}\right)  \right]  \mathbf{B}\,. \label{nr8.2}%
\end{equation}

In the case under consideration, one can construct a $\theta$-modified
pseudoclassical action $S^{\theta}$ (\textit{\`{a} la} Berezin-Marinov
\cite{BerMar77,All1}) for the nonrelativistic spinning particle. Such an
action has the form%
\begin{align}
&  S^{\theta}=\int dt\,L^{\theta},\ \ L^{\theta}=\mathbf{p}\cdot
\mathbf{\dot{q}}-\frac{1}{2m}\left(  \mathbf{p}-\frac{e}{c}\mathbf{A}\left(
\mathbf{q},t\right)  \right)  ^{2}-eA^{0}\left(  \mathbf{q}\right) \nonumber\\
&  +i\boldsymbol{\xi}\cdot\boldsymbol{\dot{\xi}}+\frac{ie}{mc}\left(
\mathbf{B}^{\theta}\cdot\left[  \boldsymbol{\xi}\times\boldsymbol{\xi}\right]
\right)  -\frac{1}{2\hslash}\dot{p}^{i}\theta^{ij}p^{j}\,, \label{s1.1}%
\end{align}
where the variables $\mathbf{q}$ and $\mathbf{p}$ describe particle spacial
movement, and Grassmann variables $\boldsymbol{\xi}=\left(  \xi^{i}%
,\ i=1,2,3\right)  $, $\left[  \xi^{i},\xi^{j}\right]  _{+}=0$, describe the
particle spin. A quantization procedure presented below results in the
nonrelativistic quantum mechanics of a spinning particle with the Hamiltonian
$\hat{H}_{\mathrm{M-Nonrel}}^{\theta}$ defined by eq. (\ref{e2}) and
commutation relations (\ref{c1}).

Treating (\ref{s1.1}) as a first order action in variables $\mathbf{q}$,
$\mathbf{p,}$ and $\boldsymbol{\xi}$, we introduce canonical momenta:%
\begin{equation}
\pi_{qi}=\frac{\partial L^{\theta}}{\partial\dot{q}^{i}}=p^{i}\,,\ \ \pi
_{pi}=\frac{\partial L^{\theta}}{\partial\dot{p}^{i}}=-\frac{1}{2\hslash
}\theta^{ij}p^{j}\,,\ \ \pi_{\xi i}=\frac{\partial_{r}L^{\theta}}{\partial
\dot{\xi}^{i}}=i\xi^{i}\,. \label{s1.2}%
\end{equation}
Equations (\ref{s1.2}) imply the primary constraints $\Phi_{a}^{\left(
1\right)  }=\left(  \Phi_{qi}^{\left(  1\right)  },\,\Phi_{pi}^{\left(
1\right)  },\,\Phi_{\xi i}^{\left(  1\right)  }\right)  $,%
\begin{equation}
\Phi_{qi}^{\left(  1\right)  }=\pi_{qi}-p^{i}\,,\ \ \Phi_{pi}^{\left(
1\right)  }=\pi_{pi}+\frac{1}{2\hslash}\theta^{ij}p^{j}\,,\ \ \Phi_{\xi
i}^{\left(  1\right)  }=\pi_{\xi i}-i\xi^{i}, \label{s1.3}%
\end{equation}
which already are second-class constraints. Constructing the Hamiltonian
$H^{\left(  1\right)  }=H^{\theta}+\lambda^{a}\Phi_{a}^{\left(  1\right)  },$
according the canonical procedure \cite{GitTy90}, we obtain%
\begin{equation}
H^{\theta}=\frac{1}{2m}\left(  \mathbf{p}-\frac{e}{c}\mathbf{A}\left(
\mathbf{q},t\right)  \right)  ^{2}+eA^{0}\left(  \mathbf{q}\right)  -\frac
{ie}{mc}\left(  \mathbf{B}^{\theta}\cdot\left[  \boldsymbol{\xi}%
\times\boldsymbol{\xi}\right]  \right)  \,.\nonumber
\end{equation}
In the case under consideration all the Lagrange multipliers $\lambda^{a}$ can
be fixed by the consistency conditions,%
\[
\dot{\Phi}_{a}^{\left(  1\right)  }=0\rightarrow\lambda^{a}=\left\{
\Phi^{\left(  1\right)  },\Phi^{\left(  1\right)  }\right\}  _{ab}%
^{-1}\left\{  \Phi_{b}^{\left(  1\right)  },H\right\}  .
\]

Performing a canonical transformation to primed variables,%
\begin{equation}
q^{i\prime}=q^{i},\ \ p^{\prime i}=p^{i},\ \ \pi_{qi}^{\prime}=\pi_{qi}%
-p^{i},\ \ \pi_{pi}^{\prime}=\pi_{pi}-q^{i}\,, \label{s1.5}%
\end{equation}
we obtain that the constraints in the new variables have the form:%
\begin{equation}
\Phi_{qi}^{\prime\left(  1\right)  }=\pi_{qi}^{\prime}=0\,,\ \ \Phi
_{pi}^{\prime\left(  1\right)  }=q^{\prime i}+\pi_{pi}^{\prime}+\frac
{1}{2\hslash}\theta^{ij}p^{\prime j}=0\,. \label{s1.6}%
\end{equation}
These constraints have a special form \cite{GitTy90}, such that variables
$\left(  q^{i\prime},\pi_{qi}^{\prime}\right)  $ can be eliminated from the
consideration and for the rest of the variables $\left(  p^{\prime i},\pi
_{pi}^{\prime}\right)  $, we have the total Hamiltonian%
\begin{align}
&  H^{\left(  1\right)  \prime}=H_{\ast}^{\theta}+\lambda_{\xi}^{i}\Phi_{\xi
i}^{\left(  1\right)  },\ \ \lambda_{\xi}^{i}=-\left\{  \Phi_{\xi}^{\left(
1\right)  },\Phi_{\xi}^{\left(  1\right)  }\right\}  _{ij}^{-1}\left\{
\Phi_{\xi j}^{\left(  1\right)  },H\right\}  \,,\nonumber\\
&  H_{\ast}^{\theta}=H^{\theta}\left(  -\pi_{pi}^{\prime}-\frac{1}{2\hslash
}\left(  \theta^{ij}p^{\prime j}\right)  ,p^{\prime};\xi\right)  \,.
\label{ham}%
\end{align}
\ 

After an additional canonical transformation $p^{i}=p^{\prime i}\,$%
,$\ \ \pi_{pi}^{\prime}=-x^{i}\,,$ i.e., $q^{i}=x^{i}-\frac{1}{2\hslash}%
\theta^{ik}p^{k}$\ , the Hamiltonian $H_{\ast}^{\theta}$ takes the form
$H^{\theta}=H^{\theta}\left(  q,\,p;\,\xi\right)  \,.$

Now, we can pass to Dirac brackets and the Hamiltonian $H^{\theta}\left(
q,\,p;\,\xi\right)  $. The only nonzero Dirac brackets between the remaining
variables are:%
\begin{align*}
\left\{  x^{i},p^{j}\right\}  _{D\left(  \Phi\right)  }  &  =\delta
^{ij}\,,\ \ \left\{  \xi^{i},\xi^{j}\right\}  _{D\left(  \Phi\right)  }%
=-\frac{i}{2}\delta^{ij}\,,\\
\left\{  q^{i},q^{j}\right\}  _{D\left(  \Phi\right)  }  &  =\frac{1}{\hslash
}\theta^{ij}\,,\ \ \left\{  q^{i},p^{j}\right\}  _{D\left(  \Phi\right)
}=\delta^{ij}\,.
\end{align*}

Proceeding to the quantization, we assign operators to classical variables
$\left(  q,p,\xi\right)  $, such that the only nonzero commutators are%
\begin{align}
&  \left[  \hat{q}^{i},\hat{q}^{j}\right]  =i\hslash\left.  \left\{
q^{i},q^{j}\right\}  _{D\left(  \Phi\right)  }\right\vert _{\eta=\hat{\eta}%
}=i\theta^{ij}\,,\nonumber\\
&  \left[  \hat{x}^{i},\hat{p}^{j}\right]  =\left[  \hat{q}^{i},\hat{p}%
^{j}\right]  =i\hslash\left.  \left\{  q^{i},p^{j}\right\}  _{D\left(
\Phi\right)  }\right\vert _{\eta=\hat{\eta}}=i\hslash\delta^{ij}\,,\nonumber\\
&  \left[  \hat{\xi}^{i},\hat{\xi}^{j}\right]  _{+}=i\hslash\left.  \left\{
\xi^{i},\xi^{j}\right\}  _{D\left(  \Phi\right)  }\right\vert _{\eta=\hat
{\eta}}=\frac{\hslash}{2}\delta^{ij}\,. \label{alg}%
\end{align}
The algebra (\ref{alg}) can be realized in a Hilbert space, whose elements
$\psi\left(  x\right)  $ are two components spinors dependent on $x$, such
that,%
\begin{equation}
\hat{x}^{i}=x^{i},\ \ \hat{p}^{i}=\mathbf{-}i\hslash\frac{\partial}{\partial
x^{i}},\ \ \hat{q}^{i}=x^{i}+\frac{i}{2}\theta^{ij}\partial_{j}\,,\ \ \hat
{\xi}^{i}=\frac{\sqrt{\hslash}}{2}\sigma^{i}\,. \label{s1.19}%
\end{equation}
It follows from (\ref{s1.19}) and (\ref{ham}) that the corresponding quantum
Hamiltonian coincides with the Hamiltonia (\ref{e2}), and the nonrelativistic
spin operator reads $\mathbf{\hat{s}}=-i\left[  \boldsymbol{\hat{\xi}}%
\times\boldsymbol{\hat{\xi}}\right]  =\left(  \hslash/2\right)
\boldsymbol{\sigma}\,$.

\subsection{Consideration in the framework of SW map}

Here we are going to obtain a $\theta$-modified Dirac equation from the SW map
applied to the action (\ref{wm0}). It is known that such a modified action is
already gauge invariant under the gauge transformations%
\[
\check{U}_{\check{\lambda}}\left(  x\right)  =\left(  e^{i\check{\lambda
}\left(  x\right)  }\right)  _{\star}=1+i\check{\lambda}\left(  x\right)
-\left(  1/2\right)  \check{\lambda}\left(  x\right)  \star\check{\lambda
}\left(  x\right)  +O\left(  \check{\lambda}^{3}\right)  ,
\]
where $\check{\lambda}\left(  x\right)  $ is the noncommutative gauge
parameter, \cite{SeiWit99}. At the first step, we take the action%

\begin{align}
&  S_{\mathrm{SW}}^{\theta}=\int d^{4}x\mathcal{L}_{\mathrm{SW}}^{\theta
}\,,\ \ \mathcal{L}_{\mathrm{SW}}^{\theta}=\overline{\check{\Psi}}\left(
x\right)  \star\left(  \gamma^{\mu}\check{P}_{\mu}-mc\right)  \star\check
{\Psi}\left(  x\right)  \,,\nonumber\\
&  \check{F}^{\mu\nu}\left(  x\right)  =\partial^{\mu}\check{A}^{\nu}\left(
x\right)  -\partial^{\nu}\check{A}^{\mu}\left(  x\right)  +\frac{ie}{\hslash
c}\left[  \check{A}^{\mu}\left(  x\right)  \overset{\star}{,}\check{A}^{\nu
}\left(  x\right)  \right]  \,,\nonumber\\
&  \left[  \check{A}^{\mu}\left(  x\right)  \overset{\star}{,}\check{A}^{\nu
}\left(  x\right)  \right]  =\check{A}^{\mu}\left(  x\right)  \star\check
{A}^{\nu}\left(  x\right)  -\check{A}^{\nu}\left(  x\right)  \star\check
{A}^{\mu}\left(  x\right)  \,,\nonumber\\
&  \left[  \check{P}_{\mu}\overset{\star}{,}\check{P}_{\nu}\right]
=-\frac{i\hslash e}{c}\check{F}^{\mu\nu}\left(  x\right)  \,, \label{sw0}%
\end{align}
where (in the first order in $\theta$) the SW fields $\check{A}_{\mu}\left(
x\right)  $ and $\check{\Psi}\left(  x\right)  $ are expressed via the
ordinary field $A_{\mu}\left(  x\right)  $ and $\Psi\left(  x\right)  $ as:%
\begin{align*}
&  \check{A}_{\mu}\left(  x\right)  =A_{\mu}\left(  x\right)  +\frac
{e}{2\hslash c}\theta^{\alpha\beta}A_{\alpha}\left(  x\right)  \left(
\partial_{\beta}A_{\mu}\left(  x\right)  +F_{\beta\mu}\left(  x\right)
\right)  +O\left(  \theta^{2}\right)  \,,\\
&  \check{\Psi}\left(  x\right)  =\Psi\left(  x\right)  +\frac{e}{2\hslash
c}\theta^{\alpha\beta}A_{\alpha}\left(  x\right)  \partial_{\beta}\Psi\left(
x\right)  +O\left(  \theta^{2}\right)  \,.
\end{align*}
see\footnote{Here the noncommutative SW fields are labeled by a
\textquotedblleft check\textquotedblright\ above.}
\cite{SeiWit99,DouNek01,GurJacPiPol01,Cai01,CarHarKosLanOka01,BicGriPopSchWul01}%
. Then we consider the star product in the first order in $\theta$ and finally
we obtain the spinor field action (e.g.
\cite{CarHarKosLanOka01,BicGriPopSchWul01}),%
\begin{equation}
\mathcal{L}_{\mathrm{SW}}^{\theta}=\bar{\Psi}\left(  x\right)  \left\{
\gamma^{\mu}\left[  \left(  1+\frac{e}{4\hslash c}\theta^{\alpha\beta
}F_{\alpha\beta}\right)  \hat{P}_{\mu}-\frac{e}{2\hslash c}\theta^{\alpha
\beta}F_{\alpha\mu}\hat{P}_{\beta}\right]  -mc\left(  1+\frac{e}{4\hslash
c}\theta^{\alpha\beta}F_{\alpha\beta}\right)  \right\}  \Psi\left(  x\right)
\,. \label{sw1}%
\end{equation}

We identify the Euler-Lagrange equation $\delta S_{\mathrm{SW}}^{\theta
}/\delta\bar{\Psi}\left(  x\right)  =0$ with the $\theta$-mod. Dirac equation
from SW map. For $\theta^{0\mu}=0,$ we obtain:%
\begin{align}
&  i\hslash\partial_{t}\Psi=\left(  \hat{\mathbb{H}}_{\mathrm{D}}+\Delta
\hat{\mathbb{H}}_{\mathrm{SW}}^{\theta}\right)  \Psi\,,\nonumber\\
&  \Delta\hat{\mathbb{H}}_{\mathrm{SW}}^{\theta}=\frac{e}{2\hslash}\left\{
\left(  \left[  \mathbf{E\times\hat{P}}\right]  \cdot\boldsymbol{\theta
}\right)  +\left(  \left[  \boldsymbol{\theta}\times\left[  \boldsymbol{\alpha
}\times\mathbf{B}\right]  \right]  \cdot\mathbf{\hat{P}}\right)  \right\}
\,,\nonumber\\
&  \mathbf{E}=\left(  E^{i}=F^{i0}\right)  \,,\ \ \mathbf{B}%
=\boldsymbol{\nabla}\times\mathbf{A}\,. \label{sw2.1}%
\end{align}
We stress that this equation is already gauge invariant under $U_{\lambda
}\left(  1\right)  $ gauge transformations.

In the same manner, as it was done in the previous subsection, we can derive
the nonrelativistic limit of the obtained equation. As a result, we obtain the
following Schr\"{o}dinger equation for a spinor $\psi\left(  x\right)  :$%
\begin{align}
&  i\hslash\partial_{t}\psi\left(  x\right)  =\hat{H}_{\mathrm{SW}}^{\theta
}\psi\left(  x\right)  \,,\nonumber\\
&  \hat{H}_{\mathrm{SW}}^{\theta}=\frac{1}{2mc^{2}}\mathcal{O}_{\mathrm{SW}%
}^{2}+eA_{0}+\frac{e}{2\hslash}\left(  \left[  \mathbf{E\times\hat{P}}\right]
\cdot\boldsymbol{\theta}\,\right)  \,,\nonumber\\
&  \mathcal{O}_{\mathrm{SW}}=c\left\{  \boldsymbol{\alpha}+\frac{e}{2\hslash
c}\left[  \boldsymbol{\theta}\times\left[  \boldsymbol{\alpha}\times
\mathbf{B}\right]  \right]  \right\}  \cdot\mathbf{\hat{P}}\,. \label{sw3}%
\end{align}
Complete operators $\mathcal{O}_{\mathrm{SW}}^{2}$\ and $\hat{H}_{\mathrm{SW}%
}^{\theta}$ (with all the terms up to the order $\left(  mc^{2}\right)  ^{-2}%
$) are presented in the Appendix.

If we restrict ourselves by external homogeneous magnetic field $\mathbf{B}%
=\left(  B^{i}\left(  t\right)  \right)  $ only, we reduce the Hamiltonian
$\hat{H}_{\mathrm{SW}}^{\theta}$ to the following form:%
\begin{align}
&  \hat{H}_{\mathrm{SW-Pauli}}^{\theta}=\frac{1}{2m}\mathbf{\hat{P}}%
^{2}+eA_{0}+\frac{e}{2\hslash}\left(  \left[  \mathbf{E\times\hat{P}}\right]
\cdot\boldsymbol{\theta}\right) \nonumber\\
&  +\frac{e}{2m\hslash c}\left(  \mathbf{\hat{P}}\cdot\left[  \mathbf{B}%
\times\left[  \mathbf{\hat{P}}\times\boldsymbol{\theta}\right]  \right]
\right)  -\boldsymbol{\hat{\mu}}\cdot\mathbf{B}_{\mathrm{SW}}^{\theta}\,,
\label{sw4}%
\end{align}
where%
\begin{equation}
\mathbf{B}_{\mathrm{SW}}^{\theta}=\left[  1+\frac{e}{\hslash c}\left(
\mathbf{B}\cdot\boldsymbol{\theta}\right)  \right]  \mathbf{B}\,. \label{B_SW}%
\end{equation}

In contrast to the nonrelativistic equation with the Hamiltonian (\ref{wm10}),
equation (\ref{sw3}), from our point of view, is a good candidate to be
considered the $\theta$-modified Pauli equation, since $\hat{H}%
_{\mathrm{SW-Pauli}}^{\theta}$ is gauge invariant. For the same reason, we can
identify now the quantity $\mathbf{B}_{\mathrm{SW}}^{\theta}$\ with a physical
($\theta$-modified) magnetic field.

It should be noted that there exists an original manner to obtain a
nonrelativistic wave equation for spinning particles from group theoretical
considerations presented in \cite{Levy-Leblond}. It should be interesting to
see if similar considerations (maybe suitable modified due to the well-known
problems with the classical space-time symmetries in noncommutative spaces,
e.g. with the Lorentz invariance in the relativistic case, see e.g.
\cite{ChaKulNisTur04}) can work in the noncommutative case.

\section{Two spins in a noncommutative space}

It is known, that a reduction of the Pauli equation to the ($0+1$)-dimensional
case, allows one to obtain the so-called spin equation (in absence of the
scalar potential, $A_{0}=0$), which describes a motion of a spatially
\textquotedblleft frozen\textquotedblright\ spin in a magnetic field, see
\cite{BagBaGL05}. The same reduction of the equation (\ref{sw4}) yields the
spin equation in the noncommutative space, or $\theta$-modified spin equation:%
\begin{equation}
i\hslash\partial_{t}\psi=-\boldsymbol{\hat{\mu}}\cdot\mathbf{B}\left(
1+\frac{e}{\hslash c}\left(  \mathbf{B}\cdot\boldsymbol{\theta}\right)
\right)  \psi\,. \label{wm12}%
\end{equation}

In the commutative space, the spin equation for two interaction spins in a
homogeneous magnetic field can be written as \cite{BagBaGL07},%
\begin{align}
&  i\hslash\partial_{t}\Psi\left(  t\right)  =\hat{H}\left(  \mathbf{B}%
_{1}\mathbf{,B}_{2}\mathbf{,}J\right)  \Psi\left(  t\right)  \,,\ \hat
{H}\left(  \mathbf{B}_{1}\mathbf{,B}_{2}\mathbf{,}J\right)  =\boldsymbol{\rho
}\cdot\mathbf{B}_{1}+\boldsymbol{\Sigma}\cdot\mathbf{B}_{2}+\frac{J}%
{2}\boldsymbol{\Sigma}\cdot\boldsymbol{\rho}\,,\nonumber\\
&  \boldsymbol{\Sigma}=I\otimes\boldsymbol{\sigma}\,,\;\boldsymbol{\rho
}=\boldsymbol{\sigma}\otimes I\,,\ \ \left(  \boldsymbol{\Sigma}%
\cdot\boldsymbol{\rho}\right)  =\boldsymbol{\sigma}\otimes\boldsymbol{\sigma
}=\sum_{i=1}^{3}\sigma_{i}\otimes\sigma_{i}\,, \label{TCS}%
\end{align}
where $I$ stand for a $2\times2$ unit matrix. The first (second) term on
$\hat{H}$ represents the Pauli interaction of the first (second) spin with the
magnetic field $\mathbf{B}_{1}\left(  t\right)  $ ($\mathbf{B}_{2}\left(
t\right)  $) and the last term a spherically symmetric interaction $J=J\left(
t\right)  $\ between the two spins (a Heisenberg interaction). In the above
expression we are absorbing the magnetic momentum of the particle on the
magnetic field $\left(  -\mu_{B}\mathbf{B}\equiv\mathbf{B}\right)  $. In this
manner a different effective field in each spin can be obtained by using
particles with different magnetic moments.

As we know from the result (\ref{sw4}), the noncommutative Pauli interaction
for a spin in a homogeneous magnetic field can be obtained by the replacement
$\mathbf{B}_{a}\left(  t\right)  \mathbf{\rightarrow B}_{a}^{\theta}\left(
t\right)  ,\ a=1,2,$ see eq. (\ref{B_SW}). So, if the fields $\mathbf{B}_{a}$
are parallel (we choose them in the $z$-direction) in the noncommutative case
the Hamiltonian (\ref{TCS})\ becomes%

\begin{align*}
&  \hat{H}^{\theta}=\frac{1}{2}\left[  \left(  \Sigma_{3}+\rho_{3}\right)
B_{+}^{\theta}-\left(  \Sigma_{3}-\rho_{3}\right)  B_{-}^{\theta}-J^{\theta
}\right]  +AJ^{\theta},\\
&  B_{\pm}^{\theta}\left(  t\right)  =B_{1}^{\theta}\left(  t\right)  \pm
B_{2}^{\theta}\left(  t\right)  \,,\ \ A=\frac{1}{2}\left[  1+\left(
\boldsymbol{\Sigma}\cdot\boldsymbol{\rho}\right)  \right]  \,,
\end{align*}
where we are supposing that the new noncommutative interaction $J^{\theta}%
$\ remains spherically symmetric. Using the techniques described in
\cite{BalGi07}, it is possible to show that the evolution operator for the
Schr\"{o}dinger equation with the above Hamiltonian has the form%
\[
U\left(  t\right)  =\left(
\begin{array}
[c]{ccc}%
f_{+}\left(  t\right)  & 0 & 0\\
0 & \hat{u}\left(  t\right)  & 0\\
0 & 0 & f_{-}\left(  t\right)
\end{array}
\right)  \,,\ \ f_{\pm}\left(  t\right)  =\exp\left[  -i\int_{0}^{t}\left(
\frac{J^{\theta}}{2}\pm B_{+}\right)  \,d\tau\right]  \,,
\]
where the $2\times2$ matrix $\hat{u}$ is the evolution operator for the
following two level system problem \cite{BagBaGL05},%
\[
i\partial_{t}\psi=\left[  \left(  \boldsymbol{\sigma}\cdot\mathbf{K}\right)
-J^{\theta}/2\right]  \psi\,,\ \ \mathbf{K}\left(  t\right)  =\left(
J^{\theta}\left(  t\right)  ,0,B_{-}^{\theta}\left(  t\right)  \right)  \,.
\]

Having this operator in the explicit form, one can calculate the
probability\ transition $P\left(  t\right)  $\ between any states. An
interesting special case is the transition between the two orthogonal EPR
states $\left\vert \Psi_{\pm}\right\rangle $,%
\begin{align*}
&  \left\vert \Psi_{\pm}\right\rangle =\frac{1}{\sqrt{2}}\left[  \left\vert
++\right\rangle \pm\left\vert --\right\rangle \right]  \ ,\ \left\vert \pm
\pm\right\rangle =\left\vert \pm\right\rangle \otimes\left\vert \pm
\right\rangle \,,\\
&  \left\vert +\right\rangle =\left(
\begin{array}
[c]{c}%
1\\
0
\end{array}
\right)  \,,\ \ \left\vert -\right\rangle =\left(
\begin{array}
[c]{c}%
0\\
1
\end{array}
\right)  \,,
\end{align*}
once, in this case, this probability does not depends on $\hat{u}$ neither on
the unknown function $J^{\theta}$,%
\[
P\left(  t\right)  =\left\vert \left\langle \Psi_{+}\right\vert U\left(
t\right)  \left\vert \Psi_{-}\right\rangle \right\vert ^{2}=\left\vert
\sin\left[  2\int_{0}^{t}B_{+}^{\theta}\,d\tau\right]  \right\vert ^{2}\,.
\]
In particular, for the two dephased fields%
\[
B_{1}=B\cos\left(  \omega t\right)  ,\ \ B_{2}=B\cos\left(  \omega
t+\phi\right)  ,\ \ \phi=\pi\left(  1+2n\right)  ,\ \ n\in%
\mathbb{N}
\,,
\]
where $B$ and $\omega$ are constants, we have a strictly dependent $\theta$
transition, which is zero unless $\theta$ is nonzero\footnote{Here, we have
restored magnetic momentum $\mu_{B}$.},%
\begin{equation}
P\left(  t\right)  =\left\vert \sin\left[  \theta\frac{2e\mu_{B}B^{2}}%
{\hslash^{2}c}\left(  t+\frac{\sin2\omega t}{2\omega}\right)  \right]
\right\vert ^{2}\,. \label{TCS1}%
\end{equation}

This result can be used to obtain an upper bound on $\theta$. For such an
estimation, we suppose that the magnetic field is strong enough but realistic
for laboratory conditions, let say $B=10\mathrm{T}$, the transition time is
$t=1$\textrm{s,} and the resolution of the experiment allows us to measure the
probability with the precision $0,05\%,\ $i.e., $P\left(  t\right)  <0.005$.
Then with these numbers, we obtain the following upper bound on $\theta$:%
\begin{equation}
P\left(  t\right)  <0.005\rightarrow\left\vert \theta\right\vert
\lesssim2.65\times10^{-30}\,\mathrm{m}^{2}. \label{bound1}%
\end{equation}
This result matches with another estimation, which could be obtained from
energy splitting of the hydrogen atom due to the space noncommutativity
\cite{AdoBalChaGitTur09}.

\section{Summary}

Starting with the two $\theta$-modified spinor field actions, the first one
obtained by a simple Moyal modification and the second one by the SW map, we
derive and discuss two different $\theta$-modified Dirac equations. Both
actions were already known before, see references above, however, the $\theta
$-modified Dirac equation from the SW map was represented for the first time.

Considering the nonrelativistic limit in both Dirac equations, we derived two
Schr\"{o}dinger equations for nonrelativistic spinning particles in the
noncommutative space. One of these equations is gauge invariant with respect
to $U\left(  1\right)  $ gauge transformations of the external electromagnetic
field and is interpreted by us as a $\theta$-modified Pauli equation.

Such an equation allows us to extract a $\theta$-modified nonrelativistic
interaction of the magnetic field with the particle magnetic moment. Using the
latter result, we construct a $\theta$-modified spin equation, which describes
a $\theta$-modified two-level system, and then a $\theta$-modified Heisenberg
model for two coupled spins placed in an external magnetic field.

A pseudoclassical model for a nonrelativistic spinning particle in the
noncommutative space is constructed. Its quantization leads to one of the
$\theta$-modified wave equation for such a particle.

In the framework of the Heisenberg model, we calculate the probability
transition between two orthogonal EPR (Einstein-Podolsky-Rosen) states for a
pair of spins in an oscillatory magnetic field and show that some of such
transitions, which are forbidden in the commutative space, are possible due to
the space noncommutativity. This allows us to estimate an upper bound on the
noncommtativity parameter.%

\section*{Acknowledgements}
We are indebted to Prof. Peter A. Horvathy for the helpful commentaries about the references.
T.C.A. thanks FAPESP; M.C.B. thanks FAPESP and D.M.G. thanks FAPESP and
CNPq for permanent support.%
%

\section*{Appendix}%

The complete nonrelativistic Hamiltonian $\hat{H}_{\mathrm{M}}^{\theta}$
(\ref{wm8}), with terms of the order $\left(  mc^{2}\right)  ^{-2}$ is%
\begin{align*}
&  \hat{H}_{\mathrm{M}}^{\theta}=\frac{1}{2mc^{2}}\mathcal{O}_{\mathrm{M}}%
^{2}+eA_{0}-\frac{e}{2\hslash}\left[  \boldsymbol{\nabla}A_{0}\times
\mathbf{\hat{p}}\right]  \cdot\boldsymbol{\theta}\\
&  -\frac{1}{8m^{2}c^{4}}\left[  \mathcal{O}_{\mathrm{M}},\left(  e\left[
\mathcal{O}_{\mathrm{M}},A_{0}\right]  +i\hslash\partial_{t}\mathcal{O}%
_{\mathrm{M}}-\frac{e}{2\hslash}\varepsilon_{ijk}\left[  \mathcal{O}%
_{\mathrm{M}},\partial_{i}A_{0}\hat{p}^{j}\right]  \theta^{k}\right)  \right]
\,,
\end{align*}
where the operators above are%
\begin{align*}
\mathcal{O}_{\mathrm{M}}^{2}  &  =c^{2}\mathbf{\hat{P}}^{2}-e\hslash c\left(
\mathbf{B}\cdot\boldsymbol{\sigma}\right)  -\frac{ec}{\hslash}\left\{  \left(
\left[  \mathbf{\hat{p}}\times\boldsymbol{\nabla}A^{i}\right]  \cdot
\boldsymbol{\theta}\right)  \hat{P}^{i}-\frac{i\hslash}{2}\left[
\mathbf{\hat{p}}\times\boldsymbol{\nabla}\left(  \mathbf{\nabla}%
\cdot\mathbf{A}\right)  \right]  \cdot\boldsymbol{\theta}\right. \\
&  -\left.  \frac{\hslash}{2}\left[  \boldsymbol{\nabla}\left(  \left[
\boldsymbol{\nabla}\times\mathbf{A}\right]  \cdot\boldsymbol{\sigma}\right)
\times\mathbf{\hat{p}}\right]  \cdot\boldsymbol{\theta}+\frac{e\hslash}%
{2c}\left[  \varepsilon_{ijk}\left(  \boldsymbol{\nabla}A^{i}\right)
\times\left(  \boldsymbol{\nabla}A^{j}\right)  \sigma^{k}\right]
\cdot\boldsymbol{\theta}\right\}  \,,\\
\left[  \mathcal{O}_{\mathrm{M}},\left[  \mathcal{O}_{\mathrm{M}}%
,A_{0}\right]  \right]   &  =-\hslash c^{2}\left\{  \hslash\boldsymbol{\nabla
}^{2}A_{0}+2\left[  \left(  \boldsymbol{\nabla}A_{0}\right)  \times
\mathbf{\hat{P}}\right]  \cdot\boldsymbol{\sigma}\right\} \\
&  -\frac{e\hslash c}{2}\left\{  \partial_{i}\left[  \left(
\boldsymbol{\nabla}A^{i}\right)  \times\left(  \boldsymbol{\nabla}%
A_{0}\right)  \right]  \cdot\boldsymbol{\theta}+\left[  \left(
\boldsymbol{\nabla}A^{i}\right)  \times\partial_{i}\left(  \boldsymbol{\nabla
}A_{0}\right)  \right]  \cdot\boldsymbol{\theta}\right. \\
&  +\left.  i\varepsilon_{ijk}\left(  \partial_{i}\left[  \left(
\boldsymbol{\nabla}A^{j}\right)  \times\left(  \boldsymbol{\nabla}%
A_{0}\right)  \right]  \cdot\boldsymbol{\theta}\right)  \sigma^{k}%
+i\varepsilon_{ijk}\left(  \left[  \left(  \boldsymbol{\nabla}A^{i}\right)
\times\partial_{j}\left(  \boldsymbol{\nabla}A_{0}\right)  \right]
\cdot\boldsymbol{\theta}\right)  \sigma^{k}\right\} \\
&  -ec\left\{  \varepsilon_{ijk}\left(  \left[  \left(  \boldsymbol{\nabla
}A^{i}\right)  \times\left(  \boldsymbol{\nabla}A_{0}\right)  \right]
\cdot\boldsymbol{\theta}\right)  \hat{P}^{j}\sigma^{k}-\varepsilon
_{ijk}\left(  \left[  \left(  \partial_{i}\mathbf{A}\right)  \times\left(
\boldsymbol{\nabla}A_{0}\right)  \right]  \cdot\boldsymbol{\sigma}\right)
\hat{p}^{j}\theta^{k}\right\}  \,,\\
\left[  \mathcal{O}_{\mathrm{M}},\partial_{t}\mathcal{O}_{\mathrm{M}}\right]
&  =e\hslash c\left\{  i\partial_{t}\left(  \boldsymbol{\nabla}\cdot
\mathbf{A}\right)  -\left[  \boldsymbol{\nabla}\times\left(  \partial
_{t}\mathbf{A}\right)  \right]  \cdot\boldsymbol{\sigma}+\frac{2i}{\hslash
}\left[  \left(  \partial_{t}\mathbf{A}\right)  \times\mathbf{\hat{P}}\right]
\cdot\boldsymbol{\sigma}\right\} \\
&  -\frac{ec}{2}\left\{  i\left[  \boldsymbol{\nabla}\left(
\boldsymbol{\nabla}\cdot\partial_{t}\mathbf{A}\right)  \times\mathbf{\hat{p}%
}\right]  \cdot\boldsymbol{\theta}-\left[  \boldsymbol{\nabla}\left(  \left[
\boldsymbol{\nabla}\times\left(  \partial_{t}\mathbf{A}\right)  \right]
\cdot\boldsymbol{\sigma}\right)  \times\mathbf{\hat{p}}\right]  \cdot
\boldsymbol{\theta}\right. \\
&  +\left.  \frac{i}{\hslash}\varepsilon_{ijk}\left(  \left[  \partial
_{t}\left(  \boldsymbol{\nabla}A^{i}\right)  \times\mathbf{\hat{p}}\right]
\cdot\boldsymbol{\theta}\right)  \hat{P}^{j}\sigma^{k}+2i\frac{e}{c}\left[
\mathbf{\nabla}\left(  \partial_{t}A^{i}\right)  \times\left(
\boldsymbol{\nabla}A^{i}\right)  \right]  \cdot\boldsymbol{\theta}\right\} \\
&  +e^{2}\varepsilon_{ijk}\left\{  \left(  \left[  \boldsymbol{\nabla}\left(
\partial_{t}A^{i}\right)  \times\left(  \boldsymbol{\nabla}A^{j}\right)
\right]  \cdot\boldsymbol{\theta}\right)  \sigma^{k}-\frac{i}{\hslash}\left(
\left[  \left(  \partial_{i}\mathbf{A}\right)  \times\left(  \partial
_{t}\mathbf{A}\right)  \right]  \cdot\boldsymbol{\sigma}\right)  \hat{p}%
^{j}\theta^{k}\right\}  \,,\\
\left[  \mathcal{O}_{\mathrm{M}},\left[  \mathcal{O}_{\mathrm{M}},E^{i}\hat
{p}^{j}\right]  \right]   &  =-\hslash^{2}c^{2}\left\{  \left(
\boldsymbol{\nabla}^{2}E^{i}\right)  \hat{p}^{j}+\frac{e}{c}\left(
\partial_{l}E^{i}\right)  \left(  \partial_{j}A^{l}\right)  +\frac{e}%
{c}\partial_{l}\left(  E^{i}\partial_{j}A^{l}\right)  \right\} \\
&  +ie\hslash c\left\{  \hslash\left[  \left(  \boldsymbol{\nabla}%
E^{i}\right)  \times\left(  \partial_{j}\mathbf{A}\right)  \right]
\cdot\boldsymbol{\sigma}-\hslash\left[  \boldsymbol{\nabla}\times\left(
E^{i}\partial_{j}\mathbf{A}\right)  \right]  \cdot\boldsymbol{\sigma}\right.
\\
&  -\left.  2i\hslash^{2}\left[  \left(  \mathbf{\nabla}E^{i}\right)
\times\left(  \partial_{j}\mathbf{A}\right)  \right]  \cdot\boldsymbol{\sigma
}+2\hslash c\left(  \left[  \left(  \mathbf{\nabla}E^{i}\right)
\times\mathbf{\hat{P}}\right]  \cdot\boldsymbol{\sigma}\right)  \hat{p}%
^{j}\right\}  \,.
\end{align*}

The complete nonrelativistic Hamiltonian $\hat{H}_{\mathrm{SW}}^{\theta}$
(\ref{sw3}), with terms of the order $\left(  mc^{2}\right)  ^{-2}$ reads%

\begin{align*}
&  \hat{H}_{\mathrm{SW}}^{\theta}=mc^{2}\gamma^{0}+eA_{0}+\frac{e}{2\hslash
}\left[  \mathbf{E\times\hat{P}}\right]  \cdot\boldsymbol{\theta}+\frac
{1}{2mc^{2}}\gamma^{0}\mathcal{O}_{\mathrm{SW}}^{2}\\
&  -\frac{1}{8m^{2}c^{4}}\left[  \mathcal{O}_{\mathrm{SW}},\left(  e\left[
\mathcal{O}_{\mathrm{SW}},A_{0}\right]  +i\hslash\partial_{t}\mathcal{O}%
_{\mathrm{SW}}+\frac{e}{2\hslash}\varepsilon_{ijk}\left[  \mathcal{O}%
_{\mathrm{SW}},E^{i}\hat{P}^{j}\right]  \right)  \theta^{k}\right]  \,,
\end{align*}
where%

\begin{align*}
\mathcal{O}_{\mathrm{SW}}^{2}  &  =\left(  c^{2}+\frac{ec}{\hslash}\left(
\boldsymbol{\theta}\cdot\mathbf{B}\right)  \right)  \left(  \mathbf{\hat{P}%
}^{2}-\frac{\hslash e}{c}\mathbf{B}\cdot\boldsymbol{\sigma}\right)  -\frac
{ec}{\hslash}\left(  \mathbf{B}\cdot\mathbf{\hat{P}}\right)  \left(
\boldsymbol{\theta}\cdot\mathbf{\hat{P}}\right) \\
&  -i\frac{ec}{2}\left\{  \boldsymbol{\nabla}\left(  \boldsymbol{\theta}%
\cdot\mathbf{B}\right)  \cdot\mathbf{\hat{P}}-\left(  \boldsymbol{\theta}%
\cdot\boldsymbol{\nabla}B^{i}\right)  \hat{P}^{i}+i\left[  \boldsymbol{\nabla
}\left(  \boldsymbol{\theta}\cdot\mathbf{B}\right)  \times\mathbf{\hat{P}%
}\right]  \cdot\boldsymbol{\sigma}\right. \\
&  -\left.  i\left(  \left[  \left(  \boldsymbol{\nabla}B^{i}\right)
\times\boldsymbol{\theta}\right]  \cdot\boldsymbol{\sigma}\right)  \hat{P}%
^{i}\right\}  \,,\\
\left[  \mathcal{O}_{\mathrm{SW}},\left[  \mathcal{O}_{\mathrm{SW}}%
,A_{0}\right]  \right]   &  =-\hslash c^{2}\left(  1+\frac{e}{\hslash
c}\boldsymbol{\theta}\cdot\mathbf{B}\right)  \left\{  \hslash
\boldsymbol{\nabla}^{2}A_{0}+2\left[  \boldsymbol{\nabla}A_{0}\times
\mathbf{\hat{P}}\right]  \cdot\boldsymbol{\sigma}\right\} \\
&  +ec\hslash\left\{  \left(  \mathbf{B}\cdot\boldsymbol{\nabla}\right)
\left(  \boldsymbol{\theta}\cdot\boldsymbol{\nabla}A_{0}\right)  -\frac{1}%
{2}\left(  \boldsymbol{\nabla}A_{0}\cdot\boldsymbol{\nabla}\left(
\boldsymbol{\theta}\cdot\mathbf{B}\right)  \right)  \right. \\
&  +\frac{1}{2}\left(  \boldsymbol{\theta}\cdot\boldsymbol{\nabla}%
\mathbf{B}\right)  \cdot\left(  \boldsymbol{\nabla}A_{0}\right)  -\frac
{1}{\hslash}\left(  \mathbf{B}\cdot\boldsymbol{\nabla}A_{0}\right)  \left[
\mathbf{\hat{P}}\times\boldsymbol{\theta}\right]  \cdot\boldsymbol{\sigma}\\
&  +\left.  \frac{1}{\hslash}\left(  \left[  \boldsymbol{\nabla}A_{0}%
\times\boldsymbol{\theta}\right]  \cdot\boldsymbol{\sigma}\right)  \left(
\mathbf{B}\cdot\mathbf{\hat{P}}\right)  \right\}  \,,\\
\left[  \mathcal{O}_{\mathrm{SW}},\mathcal{\partial}_{t}\mathcal{O}%
_{\mathrm{SW}}\right]   &  =ec\left(  1+\frac{e}{\hslash c}\left(
\boldsymbol{\theta}\cdot\mathbf{B}\right)  \right)  \left\{  i\hslash
\mathcal{\partial}_{t}\left(  \boldsymbol{\nabla}\cdot\boldsymbol{A}\right)
+2i\left[  \left(  \mathcal{\partial}_{t}\mathbf{A}\right)  \times
\mathbf{\hat{P}}\right]  \cdot\boldsymbol{\sigma}-\hslash\left(
\mathcal{\partial}_{t}\mathbf{B}\cdot\boldsymbol{\sigma}\right)  \right\} \\
&  -\frac{ie^{2}}{2\hslash}\left\{  \hslash\mathbf{B}\cdot\boldsymbol{\nabla
}\left(  \left(  \mathcal{\partial}_{t}\mathbf{A}\cdot\boldsymbol{\theta
}\right)  +\left(  \left[  \mathcal{\partial}_{t}\mathbf{A}\times
\boldsymbol{\theta}\right]  \right)  \cdot\boldsymbol{\sigma}\right)
+2\left(  \left[  \mathcal{\partial}_{t}\mathbf{A}\times\boldsymbol{\theta
}\right]  \cdot\boldsymbol{\sigma}\right)  \left(  \mathbf{B}\cdot
\mathbf{\hat{P}}\right)  \right\} \\
&  +i\frac{ec}{2}\left\{  -\boldsymbol{\nabla}\left(  \boldsymbol{\theta}%
\cdot\mathcal{\partial}_{t}\mathbf{B}\right)  \cdot\mathbf{\hat{P}}+\left(
\boldsymbol{\theta}\cdot\boldsymbol{\nabla}\partial_{t}B^{i}\right)  \hat
{P}^{i}+i\left[  \boldsymbol{\nabla}\left(  \mathcal{\partial}_{t}%
\mathbf{B}\cdot\boldsymbol{\theta}\right)  \times\mathbf{\hat{P}}\right]
\cdot\boldsymbol{\sigma}\right. \\
&  -\left.  i\left(  \left[  \boldsymbol{\nabla}\mathcal{\partial}_{t}%
B^{i}\times\boldsymbol{\theta}\right]  \cdot\boldsymbol{\sigma}\right)
\hat{P}^{i}-\frac{2}{\hslash}\left(  \left(  \partial_{t}\mathbf{B}\right)
\cdot\mathbf{\hat{P}}\right)  \left[  \mathbf{\hat{P}}\times\boldsymbol{\theta
}\right]  \cdot\boldsymbol{\sigma}\right\} \\
&  +\frac{e^{2}}{2}\left\{  i\left(  \partial_{t}\mathbf{A}\right)
\cdot\boldsymbol{\nabla}\left(  \boldsymbol{\theta}\cdot\mathbf{B}\right)
-i\boldsymbol{\theta}\cdot\boldsymbol{\nabla}\left(  \mathbf{B}\cdot
\partial_{t}\mathbf{A}\right)  -\left(  \boldsymbol{\theta}\cdot
\mathcal{\partial}_{t}\mathbf{B}\right)  \left(  \mathbf{B}\cdot
\boldsymbol{\sigma}\right)  \right. \\
&  +\left[  \boldsymbol{\nabla}\left(  \mathbf{B}\cdot\partial_{t}%
\mathbf{A}\right)  \times\boldsymbol{\theta}\right]  \cdot\boldsymbol{\sigma
}-\left[  \boldsymbol{\nabla}\times\left(  \partial_{t}\mathbf{A}\left(
\mathbf{B}\cdot\boldsymbol{\theta}\right)  \right)  \right]  \cdot
\boldsymbol{\sigma}+2\left(  \mathcal{\partial}_{t}\mathbf{B}\cdot
\boldsymbol{\theta}\right)  \left(  \mathbf{B}\cdot\boldsymbol{\sigma}\right)
\\
&  +\left.  i\left[  \left(  \partial_{t}\mathbf{B}\right)  \times
\boldsymbol{\theta}\right]  \cdot\mathbf{B}+\frac{2i}{\hslash}\left(
\mathbf{B}\cdot\partial_{t}\mathbf{A}\right)  \left[  \mathbf{\hat{P}}%
\times\boldsymbol{\theta}\right]  \cdot\boldsymbol{\sigma}\right\}  \,,\\
\left[  \mathcal{O}_{\mathrm{SW}},\left[  \mathcal{O}_{\mathrm{SW}},E^{i}%
\hat{P}^{j}\right]  \right]   &  =-\hslash^{2}c^{2}\left\{  \left(
\mathbf{\nabla}^{2}E^{i}\right)  \hat{P}^{j}-\frac{2e}{c}\varepsilon
_{jkl}B^{k}\left(  \partial_{l}E^{i}\right)  +\frac{e}{c}\varepsilon
_{jlk}E^{i}\partial_{l}B^{k}\right. \\
&  -\frac{2}{\hslash}\left(  \left[  \left(  \mathbf{\nabla}E^{i}\right)
\times\mathbf{\hat{P}}\right]  \cdot\boldsymbol{\sigma}\right)  \hat{P}%
^{j}+\frac{ie}{\hslash c}E^{i}\partial_{j}\left(  \mathbf{B}\cdot
\boldsymbol{\sigma}\right) \\
&  -i\frac{e}{c}E^{i}\sigma^{j}\left(  \mathbf{\nabla}\cdot\mathbf{B}\right)
-\frac{2e}{\hslash c}\left(  \mathbf{B}\cdot\boldsymbol{\sigma}\right)
E^{i}\hat{P}^{j}+\frac{2e}{\hslash c}E^{i}\sigma^{j}\left(  \mathbf{B}%
\cdot\mathbf{\hat{P}}\right) \\
&  +\left.  \frac{2ie}{c}\left(  \partial_{j}E^{i}\right)  \left(
\mathbf{B}\cdot\boldsymbol{\sigma}\right)  -\frac{2ie}{c}\left(
\mathbf{B}\cdot\boldsymbol{\nabla}E^{i}\right)  \sigma^{j}\right\}  \,.
\end{align*}

\end{document}